# Controllable Enhancements of Wi-Fi Signals at Desired Locations Without Extra Energy Using Programmable Metasurface

Ya Shuang, Hanting Zhao, Menglin Wei, Haoyang Li, Lianlin Li

**Abstract: We present for the first time an experimental demonstration on the energy allocation of commodity Wi-Fi signals in a programmable and inexpensive way. To that end, we design an electronically-programmable phase-binary coding metasurface, working at the 2.4GHz Wi-Fi frequency band, to manipulate dynamically and arbitrarily the spatial distribution of commodity Wi-Fi signals. Meanwhile, an efficient algorithm is developed to find the optimal coding sequence of the programmable metasurface such that the spatial energy of commodity Wi-Fi signals can be instantly controlled in a desirable way. Selected experimental results based on an IEEE 802.11n commercial Wi-Fi protocol have been provided to demonstrate the performance of the developed proof-of-concept system in enhancing the commodity Wi-Fi signals dynamically and arbitrarily. It could be expected that the proposed strategy will pave a promising way for wireless communications, future smart home, and so on.**

**Keywords: Programmable coding metasurface; spatial energy allocation; commodity Wi-Fi signals.**

## I. INTRODUCTION

Radio-frequency (RF) signals, extending human's senses in a contactless way, hold ever-increasing potentials in the modern society due to the unique capability of working in all-weather all-day and through-the-opaque conditions. With the widespread popularity of RF wireless local area networks, several advanced technologies have been proposed to achieve that was thought to be impossible by exploring everywhere wireless signals. Even moreover, in these relevant areas, there are increasing interests in exploring a *device-free* approach due to the elegant property of the low hardware cost and easy implementation. Recently, the commodity Wi-Fi signals can be manipulated to monitor human activities in our daily lives, such as, to "hear" what people are talking [8], to "see" what people are doing [9-11], and so on, whatever people are in a line-of sight (LoS) or non-line-of-sight (NLoS) scenario. For instance, Wang *et al* invented a device WiHear which enables hearing-impaired people to "hear" what other people are talking without wearing any acoustic hearing aid device [8], which is realized by detecting and analyzing the commodity Wi-Fi signals bounced off the human mouth of interest. Similarly, Ali *et al* developed a WiKey recognition system to do the keystroke recognition by processing Wi-Fi signals reflected from the human keyboarding fingers [9]. More recently, researchers have demonstrated that commodity Wi-Fi signals support three-dimensional (3D) holography imaging [12], regardless of the specimen being or not being behind visually opaque obstacles [13]. Nonetheless, we here would like to emphasize that these aforementioned technologies heavily rely on the use of a large-scale array of sensors, which is costly prohibitive and is hardly deployed in many practical cases.

Now, a natural question arises up: *Can the commodity Wi-Fi signals be arbitrarily enhanced at the desired locations without consuming extra energy in a programmable and inexpensive way?* The answer is positive, especially with rapid development of programmable metasurfaces. Programmable metasurfaces, which are composed of an array of electronically-controllable digital meta-atoms, have been demonstrated to be powerful in arbitrarily and dynamically manipulating electromagnetic (EM) wavefronts [22]. Until now, various versatile devices based on programmable coding metasurfaces have been invented, for instance, beam scanner [18], dynamic metasurface holograms [19-20], machine-learning imager [21]. Although these techniques work in an active manner in the sense that cooperative EM wavefronts are actively manipulated by using the programmable coding metasurface, it can be faithfully expected that such metasurface may provide an opportunity to tailor ambient Wi-Fi signals in a desirable and dynamic way.

In this work, we focus primarily on the spatial energy allocation of commodity Wi-Fi signals without consuming extra energy in a programmable and inexpensive way. To that end, we design a specialized programmable coding metasurface working at around 2.4GHz, along with an efficient optimization algorithm for instantly adjusting the coding sequence of programmable metasurface. Here, we would like to emphasize that our method is remarkably different from conventional device-oriented and retransmission-oriented approaches. As opposed to these approaches which rely on the expenditure of extra energy, our method can realize the enhancement of the commodity Wi-Fi signals by relocating the spatial energy distribution of Wi-Fi signals without consuming extra energy.

The remaining of this paper is organized as follows. In Section II, we will elaborate on design, fabrication and test of phase-binary programmable coding metasurface working at around 2.4GHz. Afterwards, we will elaborate on an efficient algorithm to achieve the optimal coding sequence such that the desirable spatial energy distribution of ambient Wi-Fi signals can be obtained. Later on, selected experimental results are provided to demonstrate the performance of the proposed methodology in controlling the spatial energy of ambient Wi-Fi signals. These experiments are performed using an IEEE 802.11n commercial wireless device (off-the-shelf) working at the $7^{th}$ channel (2.442 GHz) with bandwidth of 20 MHz. Finally, some concluding remarks are summarized in Section V.

## II. DESIGN OF PROGRAMMABLE METASURFACE

We start our discussions by elaborating on a phase-binary programmable coding metasurface working at around 2.4GHz. The designed programmable coding metasurface is composed of 24 ×24 binary-phase electronically-controllable meta atoms. The design of such meta-atoms is detailed in Subsection II.A, while the fabrication and test of the programmable metasurface are discussed in Subsection II. B.

### *A. Design of Binary-Phase Electronically-Controllable Meta-Atoms*

The sketch map of proposed binary-phase electronically-controllable meta-atom is illustrated in Figs. 1(a) and (b). The meta atom has a sandwich configuration, which is composed of a square patch and a metal ground layer spaced by the $F_4$BM substrate with dielectric constant of 2.55 and loss tangent of 0.0015. A $90^o$-bending strip near the edge of square patch will be connected to the ground plane through a metal via. A commercial PIN diode(SMP1345-079LF) with the low insertion loss (≤0.2dB) and high isolation (≥13dB) in 2.4GHz Wi-Fi frequency band has been introduced at the gap between the square path and $90^o$-bending strip. When the PIN diode is switched from “ON” (or ‘OFF’) to “OFF” (or ‘ON’), the meta-atom will experience the reflection phase difference of $180^o$ when it is illuminated by a *x*-polarized plane wave. Moreover, the entire length of the bending strip needs to be carefully adjusted such that the maximum phase difference, ideally $180^o$, when the PIN diode works at two different status of “ON” and “OFF”, can be achieved. Moreover, in order to apply appropriate external DC voltage and associated feeding lines for the PIN diode, another layer of substrate (FR-4) with thickness of 0.7 mm has been mounted below the ground plane. At this layer, a thin biasing line, connected to the other side of the patch, is attached to the control signal line under the FR-4 layer through a metal via, without connecting to the ground plane. To achieve the good isolation between the RF and DC signals, as marked in Fig. 1(a), we introduce an inductor (TDK MLK1005S33NJT000) with inductance of L = 33nH and the self-resonance frequency higher than 3.5GHz. Moreover, a current limiting resistance has also been utilized to guarantee as lower power consumption as possible.

We have performed a set of simulations using a commercial EM simulator, CST Microwave Studio 2017, to investigate the performance of developed electronically-controllable binary-phase meta-atom. In our simulations, the unit-cell boundary has been used. In addition, the PIN diode is modelled as an equivalent lumped L-R series circuit when it works at the ON status, while being as the lumped L-C series circuit at OFF, as shown in Table I. Other optimized parameters of the proposed meat atom are set as follows: the size of top square patch is 37mm by 37mm, the width and total length of $90^o$ -bending strip are 0.8mm and 14.6mm respectively, and thickness of substrate $F_4$BM and substrate FR-4 are 1.5mm and 0.7mm, respectively.

TABLE I
EQUIVALENT LUMPED CIRCUIT PARAMETERS OF THE PIN DIODE

| Parameters | “ON” | “OFF” |
|---|---|---|
| R | 2.0Ω | - |
| L | 0.7nH | 0.7nH |
| C | - | 1.8pF |

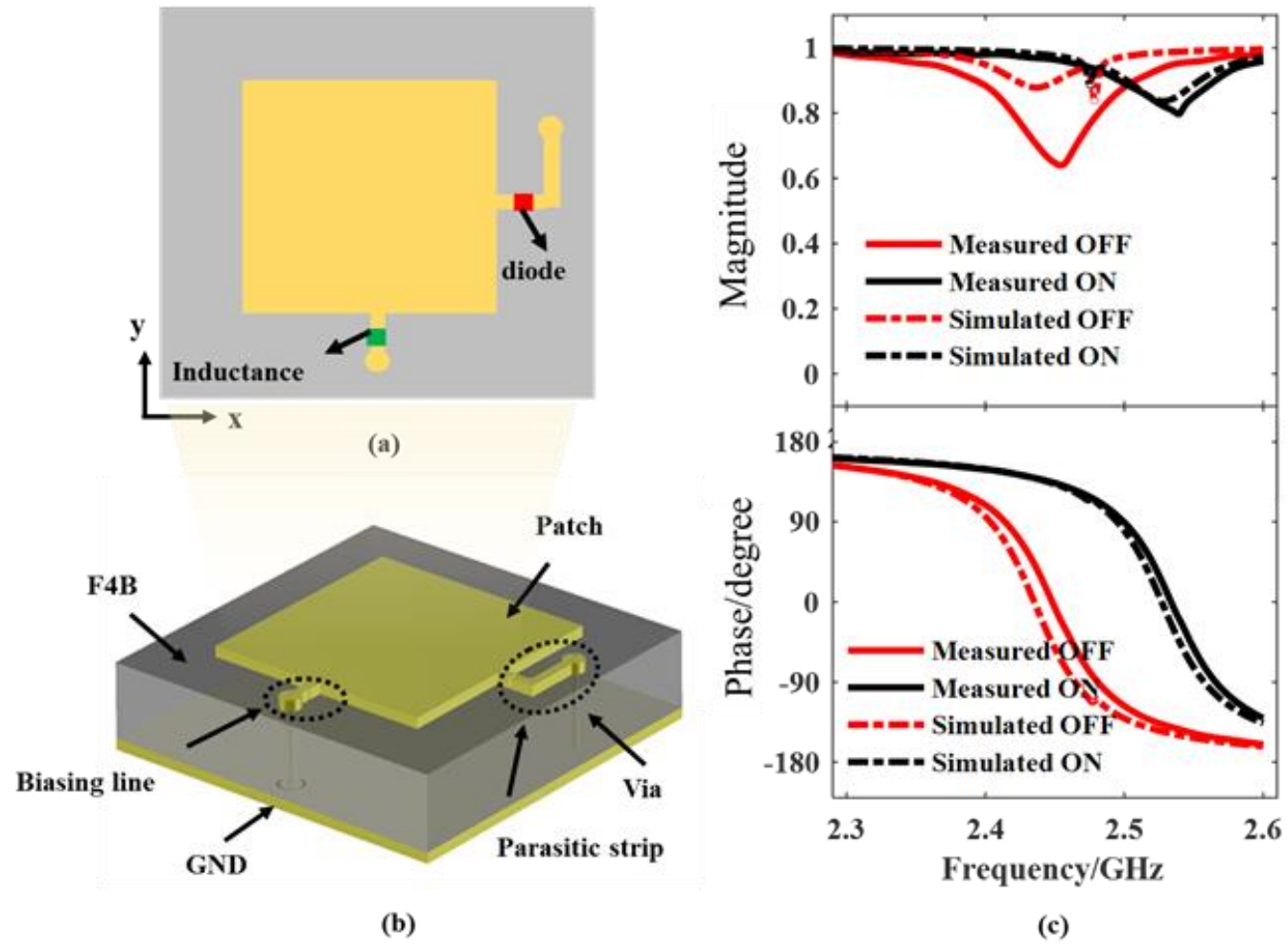

Fig. 1. The geometrical map and its properties of the proposed meta atom. (a) Top view of the meta atom. (b) Perspective view of the meta atom. (c) Simulated and experimental amplitude-frequency and phase-frequency responses of the meta atom.

Simulated and measured results of the proposed binary-phase meta-atom are plotted in Figs. 1(c). Overall, both the simulated phase-frequency response and amplitude-frequency response are in well agreements with corresponding measured results, except that the amplitude-frequency response has some discrepancy when the PIN works at OFF. Such discrepancy is probably due to the discrepancy of the PIN diode model at the OFF state, i.e., the reverse resistance should be considered. Additionally, the uncertainty of soldering and assembling errors may cause some measurement errors. Additionally, it can be readily observed from Fig. 1(c) that regardless of the operational states of PIN diode, the simulated reflection amplitudes are larger than 0.8, and the frequency bandwidth within the phase difference 180°±20°is about 55 MHz (2.401-2.456 GHz), covering 8 channels of commodity Wi-Fi Frequency bands.

### *B. Fabrication and Test of Programmable Coding Metasurface*

Figures 2(a) and 2(b) report the front view and back view of the designed programmable metasurface, respectively. The metasurface is fabricated using standard printed circuit board (PCB) technology. Moreover, the zoomed version of the meta atom is inserted in Fig. 2(a), and the FPGA-based micro control unit (MCU) is inserted in Fig. 2(b). The whole metasurface is composed of $3 \times 3$ metasurface panels, and each panel is composed of $8 \times 8$ electronically-controllable meta atoms. Then, the whole programmable coding metasurface has 24 × 24 independently controlled programmable meta atoms, and thus has the size of 1.296m×1.296m.

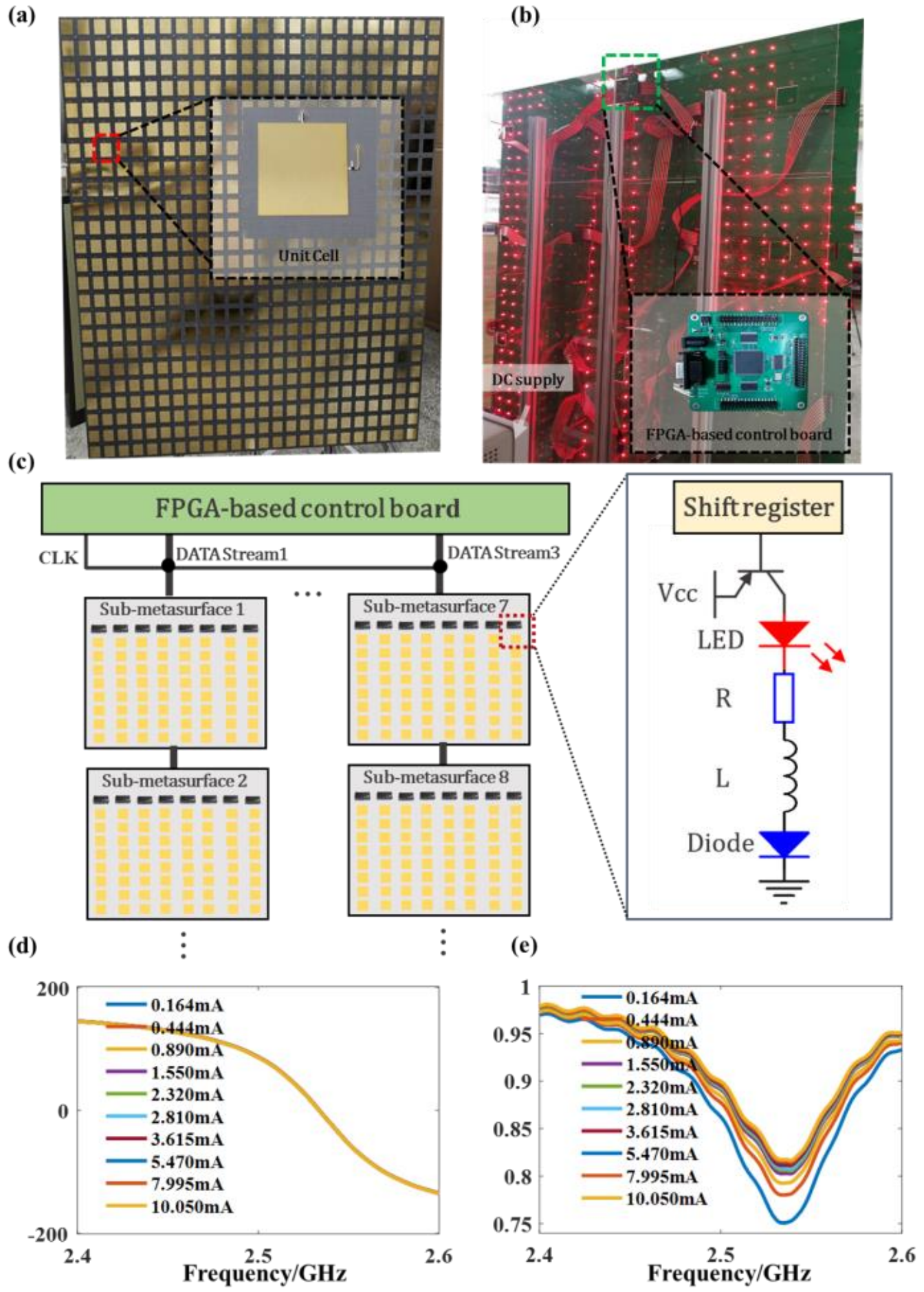

Fig. 2. The fabricated metasurface prototype and its properties. (a) The front of the metasurface. (b) The back of the metasurface. (c) The control architecture of the FPGA board and zoomed version of logical circuit on metasurface panel. (d) The phase responses of the meta atom at ON status with varying drive current. (e) The magnitude responses of the meta atom at ON status with varying drive current.

To achieve the real-time and flexible control of PIN diodes soldered on programmable coding metasurface, a dedicated MCU with size of 90mm×90mm has been designed and assembled at the upper rear of designed programmable coding metasurface (see Fig. 2(b)). This MCU is connected, in parallel, with three metasuface panels through three 1.0m-length winding wires, and each of them is connected, in series, with another two metasurface panels, as illustrated in Fig. 2(c). The MCU is in charge of dispatching all commands sent from a master computer subject to one common clock (CLK) signal. The CLK rate adopted is 50MHz, and the ideal switching time of PIN diodes in principle is 10us. One FPGA chip (Cyclone@ IV EP4CE10E22I7) is used to distribute all commands to all PIN diodes. Each metasurface panel is equipped with eight 8-bit shift registers (SN74LV595APW), and eight PIN diodes will be sequentially controlled. Consequently, the MCU will send the commands over 24 independent branch channels, leading to the almost real-time manipulation of 576 PIN diodes. Moreover, 576 red-color LEDs are employed to indicate the status of 576 PIN diodes, in particular, to indicate clearly whether the PIN diode works well or not. Here, we would like to say that the proposed control strategy can be readily extended for more PIN diodes by concatenating more metasurface panels, allowing adjustable rearrangement of metasurface panels to meet various needs.

Besides above issues, there are two another critical issues in designing the large-scale programmable coding metasurface. The first is about the insufficient driving capability of shift registers chips subject to the limited energy expenditure available. To address this issue, we introduce an external biasing voltage for each PIN diode by a BJT serving as a switch to empower the driving capability of shift registers chips, as figured out in Fig. 2(c). The second issue is about the limited energy budget available. To reduce as much power consumption as possible, for each PIN diode, a current limiting resistance is adopted to adjust its drive current. In order to determine the optimal resistance value, we have performed a set of experiments using so-called waveguide method [24] to investigate the relationship between the power consumption and EM response characteristic of the meta-atom at ON status. Two important conclusions can be drawn immediately from Fig. 2(d)-(e). First, the EM response of the designed meta-atom at the ON status remains almost unchanged as the drive current of PIN diode is larger than 1.550MA. Second, it is enough to drive a LED with satisfactory visual effect using a small current as small as 1mA. Based on above observations, for each PIN diode, we use a 680Ω resistance to limit its driving current to 3.38mA at 0.85V. Then, the total power dissipated by all PIN diodes at ON status is about 1.6w, which is much lower than that used in [25].

## III. Optimization Algorithm of Coding Sequence of Metasurface

As frequently pointed out previously, in order to manipulate ambient Wi-Fi signals in a desirable way, one critical issue is to find the matching control sequence of metasurface. As such, the status of PIN diode can be flexibly controlled such that the EM response of meta atoms to the illumination of Wi-Fi signals can be dynamically manipulated. As we known, it is a typical NP-hard combinatorial optimization problem. Although many efforts have been made to optimize the coding sequence of meta atoms, most of them focus on tailoring the far-field radiation sequence of programmable meatsurface, such as an irregular far-field radiation [18], the RCS reduction in broadband and broad-angle [23], and so on. Here, we focus on the reallocation of spatial energy of ambient Wi-Fi signals from metasurface, which is not only limited to the manipulation of far-field radiation. To this end, we need to model the EM scattering of binary-phase meta atoms with more accuracy than that used in the far-field radiation prediction, as detailed in Subsection III. A. Afterwards, based on such model, we employ the extended G-S algorithm to find the at least suboptimal coding sequence of developed programmable metasurface, as discussed in Subsection III.B.

### *A. EM Scattering Model of Binary-Phase Electronically-Controllable Metasurface.*

To illustrate our methodology, we consider a relatively simple scenario as shown in Fig. 3, where the programmable coding metasurface, composed of N × M reflection-type phase-binary electronically controllable meta atoms, is exposed to the illumination of Wi-Fi signals emitted from a wireless device. To achieve the spatial energy distribution of Wi-Fi signals with enough accuracy when the programmable coding metasurface exists, we made two-aspect efforts as follows. First, the radiation field of meta atoms has been modeled in terms of the well-known Huygens' principle, by which more realistic EM interaction between the meta atom and Wi-Fi signals can be taken into account. Second, the total scattering field from the whole coding metasurface aperture is synthesized according to the superposition principle.

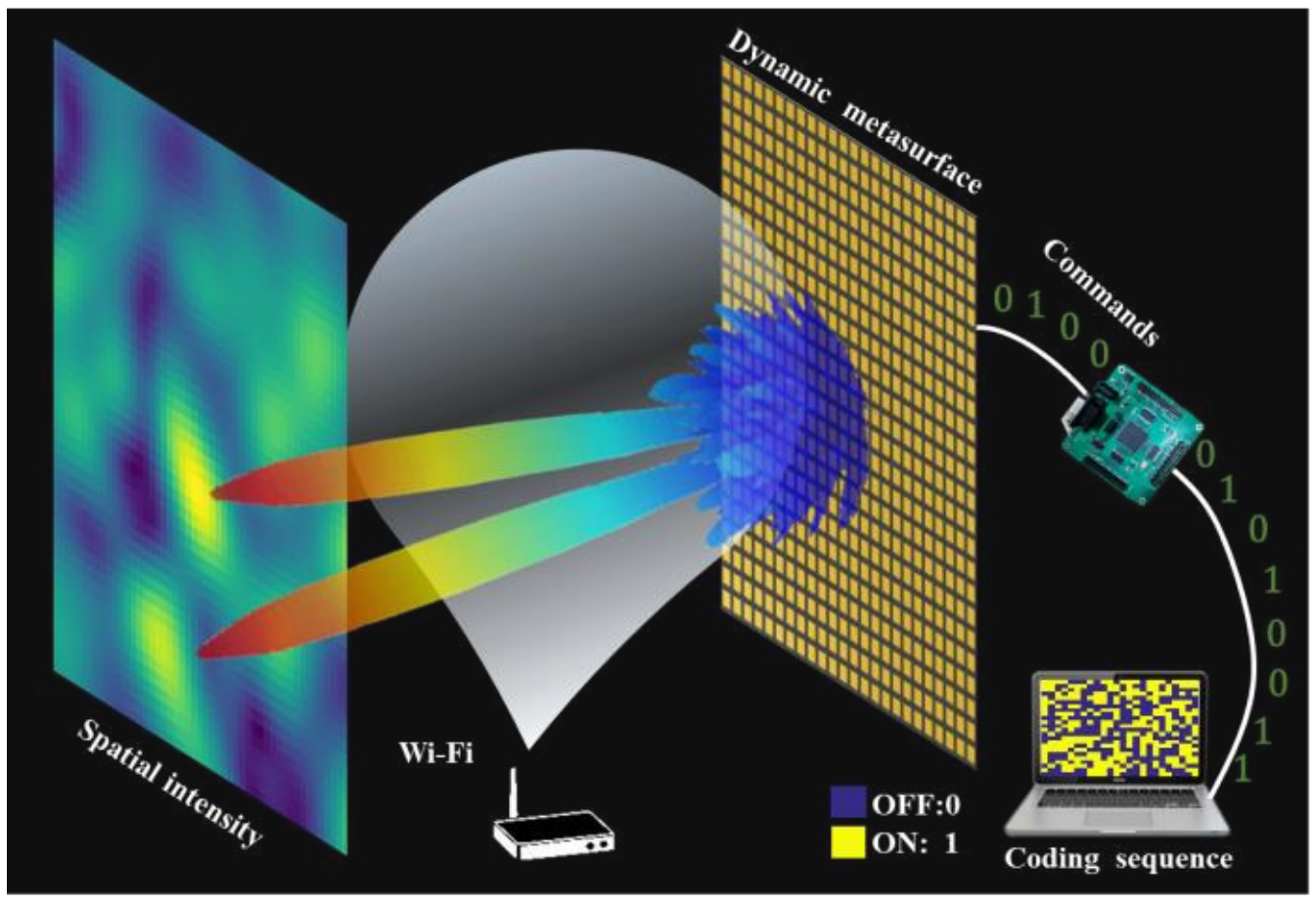

Fig.3. Schematic map of the simulation scenario.

First, we focus on the prediction of the EM responses of meta atoms. In light of the well-known Huygens' principle, we know that the EM response of meta-atoms can be accurately obtained once its induced equivalent current is obtained. Taking this fact into account, we start by evenly dividing the meta atom into $P\times Q$ grids with subwavelength scale, and then the induced current over each grid is approximated to be uniform. The relationship between the scattering field $\mathbf{E}$ of each meta atom and the induced equivalent current $\mathbf{J}$ can be represented as following, i.e.,

$$\mathbf{E}(\mathbf{r}) = \sum_{px=1}^{P}\sum_{py=1}^{Q} i\omega\mu \int_{\Delta} \bar{\mathbf{G}}\left(\mathbf{r},\mathbf{r}'_{px,py}+\delta\mathbf{r}'\right)\cdot\mathbf{J}\left(\mathbf{r}'_{px,py}+\delta\mathbf{r}'\right) d\delta\mathbf{r}' \quad (1)$$

where $\mathbf{r}'_{px,py}$ denotes the central coordinate of the $(\mathrm{px},py)^{th}$ grid of the meta atom, $\mathbf{r}$ denotes the observation position. Moreover, $\bar{\mathbf{G}}(\mathbf{r},\mathbf{r}') = \left[\bar{I}+\frac{\nabla\nabla}{k_0^2}\right]\frac{e^{jk_0|\mathbf{r}-\mathbf{r}'|}}{4\pi|\mathbf{r}-\mathbf{r}'|}$ is the Dyadic Green's function in free space, and $k_0$ is the wavenumber in vacuum. The integration in Eq. (1) is implemented over a single grid of the meta atom, whose area is Δ. When the grid area tends to 0, i.e. Δ→ 0, the scattering field $\mathbf{E}$ becomes

$$i\omega\mu\Delta\sum_{px=1}^{P}\sum_{py=1}^{Q}\bar{\mathbf{G}}\left(\mathbf{r},\mathbf{r}'_{px,py}\right)\cdot\mathbf{J}\left(\mathbf{r}'_{px,py}\right) = i\omega\mu\Delta\sum_{px=1}^{P}\sum_{py=1}^{Q}\begin{bmatrix}G^{xx}_{\mathbf{r},\mathbf{r}'_{px,py}} & G^{xy}_{\mathbf{r},\mathbf{r}'_{px,py}}\\ G^{yx}_{\mathbf{r},\mathbf{r}'_{px,py}} & G^{yy}_{\mathbf{r},\mathbf{r}'_{px,py}}\end{bmatrix}\begin{bmatrix}J^{x}_{\mathbf{r}'_{px,py}}\\ J^{y}_{\mathbf{r}'_{px,py}}\end{bmatrix},$$

where the small z component of the induced current has been ignored. For numerical implementation, Eq. (1) can be reformulated in a compact form, i.e.,

$$\mathbf{E} = i\omega\mu\Delta\begin{bmatrix}G^{xx}_{\mathbf{r},\mathbf{r}'_{11}} & G^{xy}_{\mathbf{r},\mathbf{r}'_{11}} & G^{xx}_{\mathbf{r},\mathbf{r}'_{12}} & G^{xy}_{\mathbf{r},\mathbf{r}'_{12}} & \cdots & G^{xx}_{\mathbf{r},\mathbf{r}'_{PQ}} & G^{xy}_{\mathbf{r},\mathbf{r}'_{PQ}}\\ G^{yx}_{\mathbf{r},\mathbf{r}'_{11}} & G^{yy}_{\mathbf{r},\mathbf{r}'_{11}} & G^{yx}_{\mathbf{r},\mathbf{r}'_{12}} & G^{yy}_{\mathbf{r},\mathbf{r}'_{12}} & & G^{yx}_{\mathbf{r},\mathbf{r}'_{PQ}} & G^{yy}_{\mathbf{r},\mathbf{r}'_{PQ}}\end{bmatrix}\begin{bmatrix}J^{x}_{\mathbf{r}'_{11}}\\ J^{y}_{\mathbf{r}'_{11}}\\ J^{x}_{\mathbf{r}'_{12}}\\ J^{y}_{\mathbf{r}'_{12}}\\ \vdots\\ J^{x}_{\mathbf{r}'_{PQ}}\\ J^{y}_{\mathbf{r}'_{PQ}}\end{bmatrix}$$

$$= \mathbf{AJ} = \begin{bmatrix}E^{x}_{\mathbf{r}}\\ E^{y}_{\mathbf{r}}\end{bmatrix} \quad (2)$$

which is exactly that one used in [19]. Here, $\mathbf{r}'$ goes over the central coordinate of all divided grids, $\mathbf{A}$ is a mapping matrix with entries coming from the Dyadic Green's function, and the induced current of the meta atom can be organized into a $2\times P\times Q$ column vector $\mathbf{J}$. Then, a model that a meta atom either at ON or OFF status illuminated by a plane wave is calculated through numerical simulation, and the resultant scattering field is collected at a distance far enough from the meta atom, arranged into the column vector $\mathbf{E}$. Now, the least-square method can be used to retrieve the induced equivalent current $\mathbf{J}$, namely,

$$\mathbf{J} = \left(\mathbf{A}'\mathbf{A}+\gamma\mathbf{I}\right)^{+}\mathbf{A}'\mathbf{E} \quad (3)$$

where γ denotes an artificial regularization parameter, $\mathbf{I}$ means the unit matrix, and + presents the matrix pseudo inverse. In our specific implementations, the surface of the meta atom is uniformly divided to 10×10 square grids with area of 0.044λ×0.044λ, and the scattering field $\mathbf{E}$ is obtained in an observation plane 0.5λ away from the meta-atom which is illuminated by a plane wave with intensity of 1V/m. The equivalent induced current of the meta-atom at 2.442GHz can be calculated according to Eq. (3), where $\gamma = 10^{10}$ is used.

Second, we now synthesize the EM field induced from the whole programmable coding metasurface in terms of the so-called superposition principle. For simplicity, the electrical field from a single Wi-Fi router is considered, which can be approximated as a typical spherical wave, i.e.,

$$E^{in} = \left|E^{in}\right|\exp(j\varphi_{in}) = C_0\cos(\theta_f)\frac{\exp(jk_0r_f)}{r_f} \quad (4)$$

where $C_0$ is a calibration constant, $\theta_f$ and $r_f$ denote the elevation angle and the observation distance in the spherical coordinate system centered at the Wi-Fi router, respectively. Here, the EM interaction among different meta-atoms has been ignored due to the relatively big distance among meta-atoms of being around the half operational wavelength. As a consequence, the co-polarized electrical field scattered from the whole metasurface aperture illuminated by a Wi-Fi router can be expressed as following, i.e.,

$$\mathbf{E}_s(\mathbf{r}) = \sum_{n_x=1}^{N}\sum_{n_y=1}^{M}\mathbf{A}(\mathbf{r},\mathbf{r}_{n_x,n_y})\,\mathbf{\Lambda}_{nx,ny}\mathbf{J}_{n_x,n_y}^{ON/OFF} \quad (5)$$

Herein, $\mathbf{\Lambda}_{nx,ny} = \begin{bmatrix} E_{n_x,n_y}^{in} & 0 \\ 0 & E_{n_x,n_y}^{in} \end{bmatrix}$, and the double summation involved in Eq.(5) is performed over all meta atoms of metasurface, where $(n_x, n_y)$ denotes the running indices of all meta-atoms along the $x$ and $y$ directions, $E_{n_x,n_y}^{in}$ and $\mathbf{J}_{n_x,n_y}^{ON/OFF}$ denote the incident Wi-Fi wave and the induced current on the $(n_x, n_y)^{th}$ meta atom, respectively. The flowchart of the proposed algorithm procedure has been provided in Fig. 4(a). Here, we would like to say that under the same computation configuration, the method outlined above will cost only a few minutes, however, it will take tens of hours by using the CST microwave studio package in a typical personal computer.

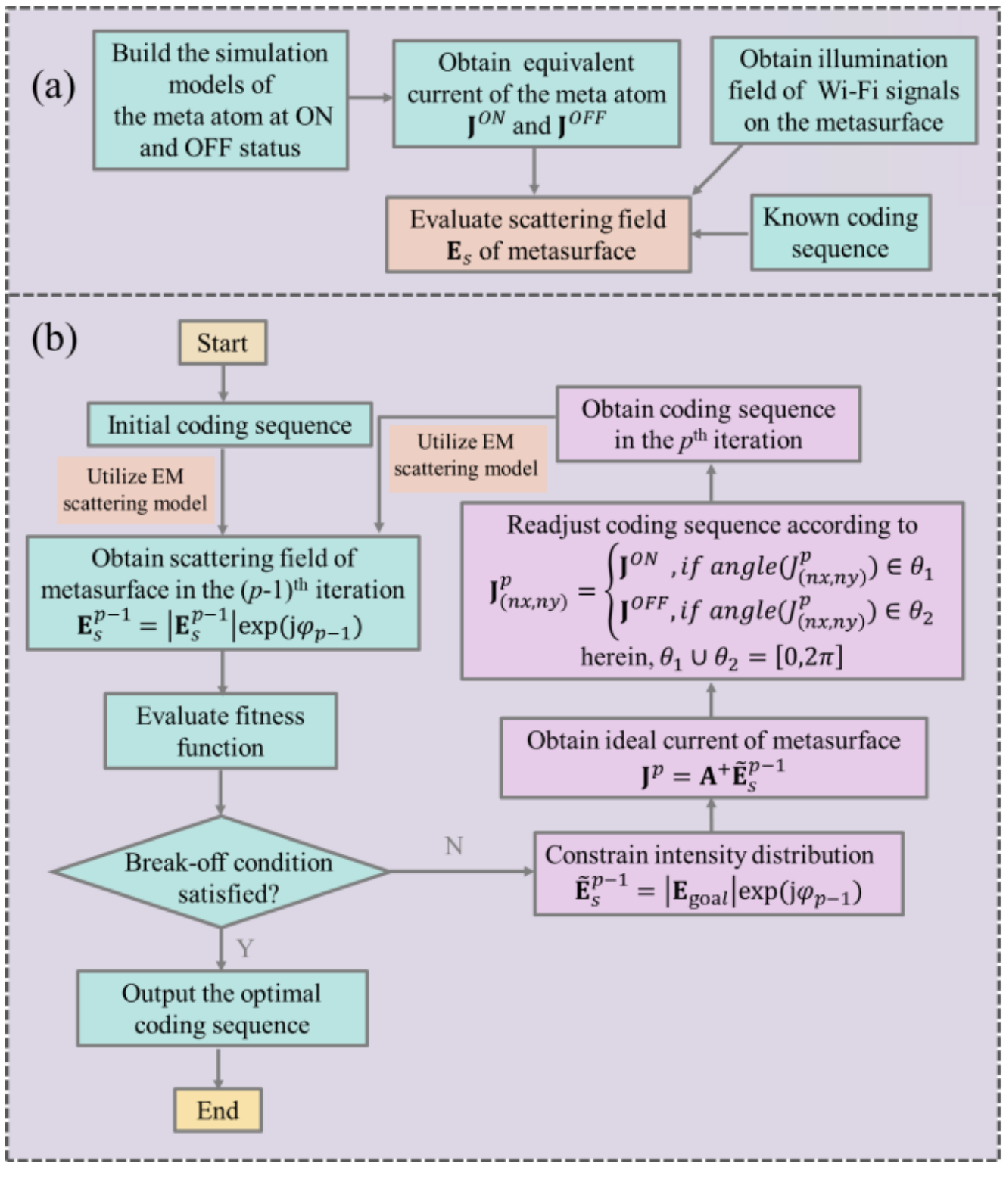

Fig. 4. The flowchart of the proposed optimization algorithm. (a) EM scattering field model of the binary-phase electronically-controllable metasurface illuminated by commodity Wi-Fi signals. (b) The modified G-S algorithm for finding the optimal coding sequence of the programmable coding metasurface.

### *B. Modified G-S Algorithm for Finding the Optimal Coding Sequence of Metasurface*

Now, the modified Gerchberg-Saxton (G-S) algorithm can be applied to find the optimal control sequence of metasuface such that the desired distribution of spatial energy of ambient Wi-Fi signals can be achieved. Mathematically, the G-S algorithm is performed to minimize the fitness function in Eq. (6), i.e.,

$$Cost = \sum_{(x,y)\in D}\sum(\mathbf{E}_{goal}(x,y) - |\mathbf{E}_s^{norm}(x,y)|)^2 \quad (6)$$

$$\mathbf{E}_{goal}(x,y) = \begin{cases}1, & (x,y) \in D_{desired} \\ 0, & (x,y) \notin D_{desired}\ \text{, and }, (x,y) \in D\end{cases} \quad (7)$$

where $\mathbf{E}_s^{norm}(x,y)$ represents the normalized scattering field from the whole metasurface, $\mathbf{E}_{goal}(x,y)$ represents the desired spatial intensity distribution in an observation plane $D$. The complete flowchart of G-S algorithm procedure has been summarized in Fig. 4(b). Apparently, such solution strategy behaves in an iterative way. As for its initial guess, the meta atoms of programmable coding metasurface are randomly set to be ON or OFF status. After a few tens of iterations in several seconds, the stable convergence can be achieved. It is stressed that different phase range $\theta_1$ and $\theta_2$ lead to different optimization results when readjusting coding sequence according to the phase of induced current of meta atoms. Therefore, it is necessary to scan the phase range to obtain better results. Besides, it is noted that the G-S algorithm is local optimization method, which usually traps into a local minimum. It is worth mentioning that a global optimization algorithm such as particle swarm optimization algorithm and genetic algorithm can be further employed to improve the output result by starting with the initial arrangement obtained by using the G-S algorithm. We will leave this discussion for the future investigation.

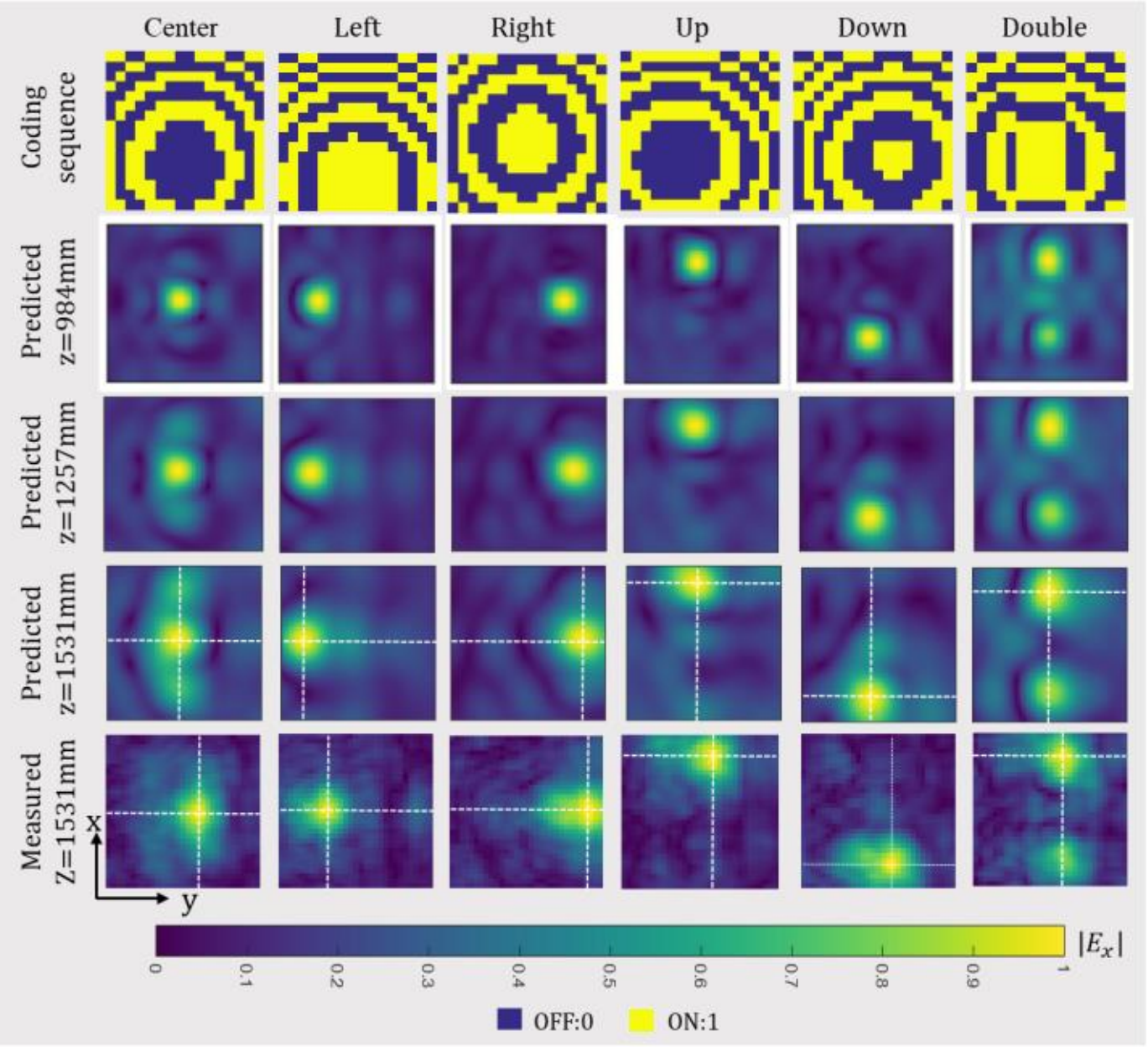

Fig. 5. Results comparison between prediction and measurement. The first row are the coding sequences. The second to fourth rows are the optimized normalized spatial intensity distributions predicted respectively at $z$=0.984m, $z$=1.257m, and $z$=1.531m using the proposed method. The fifth row are the measured normalized spatial intensity distributions at $z$=1.531m using the near-field scanning technology.

### *C. Near-field scanning of programmable coding metasurface*

In this subsection, a set of simulation and experimental results of enhancing signal intensity at desired locations have been provided to illustrate the performance of proposed methods in section III.A and B. To that end, Wi-Fi signals are mimicked by using a vector network analyzer (VNA, Agilent E5071C), and so-called near-field scanning technique is utilized to obtain the spatial distribution of Wi-Fi when the programmable coding metasurface is presented. Additionally, a Wi-Fi transmitting antenna is located at the location of B (0, -0.291m, 0.8m). The VNA is connected with the Wi-Fi antenna and an open-ended waveguide probe to scan the near-field distribution by measuring the transmission coefficients S12 in an observation plane. The scanning area is a 0.945m×0.945m square with sampling space of 0.0315m. We consider six different coding sequences of programmable metasurface, as shown in the top row of Fig. 5. Accordingly, the simulated spatial distributions of mimicked Wi-Fi signals at three observation planes in the distance of $z$=0.984m, 1.257m, and 1.531m away from the programmable coding metasurface, have been reported in the second to fourth rows of Fig. 5, respectively. For comparison, the experimental results at the observation plane of $z$=1.531m are also provided in the bottom row of Fig. 5. Note that in these figures, the electrical field distributions have been normalized by their own maximums. It is clear that the predicted intensity allocations for all given coding sequences are gradually becoming distorted with the growth of observation plane distance, which is sensible since the coding sequences have been particularly optimized at the given plane $z$=0.984m for the aforementioned desired field distributions. Better performance of intensity allocation at different observation planes will be achieved if adaptive optimizations are performed according to the position of observation distance. The experimental results at $z$=1.531m plane are obtained, which shows well agreement with the theoretical results at the identical plane and validate the effectiveness of the proposed method.

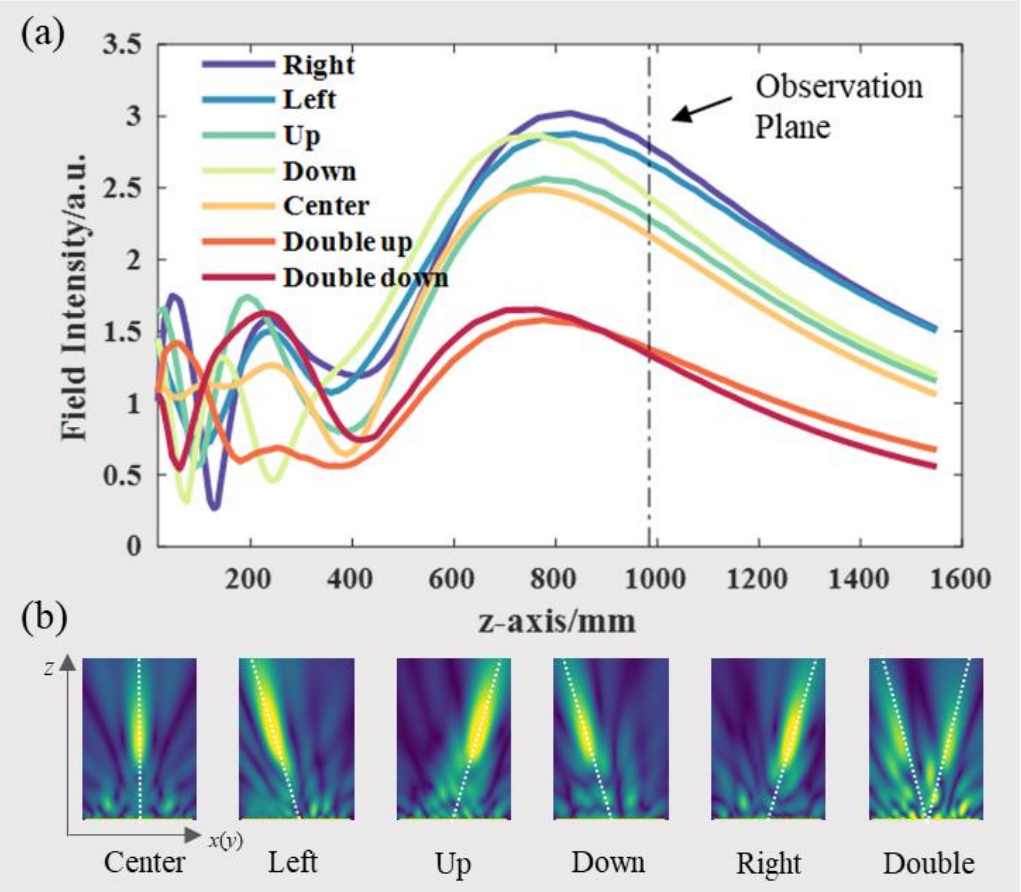

Fig. 6. The relationships between the simulated field intensity along radial direction and the distance to the metasurface along $z$-axis for six different cases, i.e., “Left”, “Right”, “Up”, “Down”, “Center”, “Double”.

Here, we will also clarify the superiority of our developed method that it can be implemented not only in far field region, but also in near filed region. The simulated field distribution in longitudinal plane (*xoz* plane or *yoz* plane) for the aforementioned six cases are provided in Fig. 6(b). And, the relationship between the simulated field intensity along radial direction (represented by a white dotted line in Fig. 6(b)) and the distance to metasurface along $z$-axis is presented in Fig. 6(a). It is apparent that the field intensity has been enhanced significantly around the designated observation distance ($z$=0.984mm), while it decreases rapidly as the distance along z-axis continues to grow. Since the observation plane belongs to near field region of the metasurface, it effectively demonstrates that the proposed method works well in near-field region.

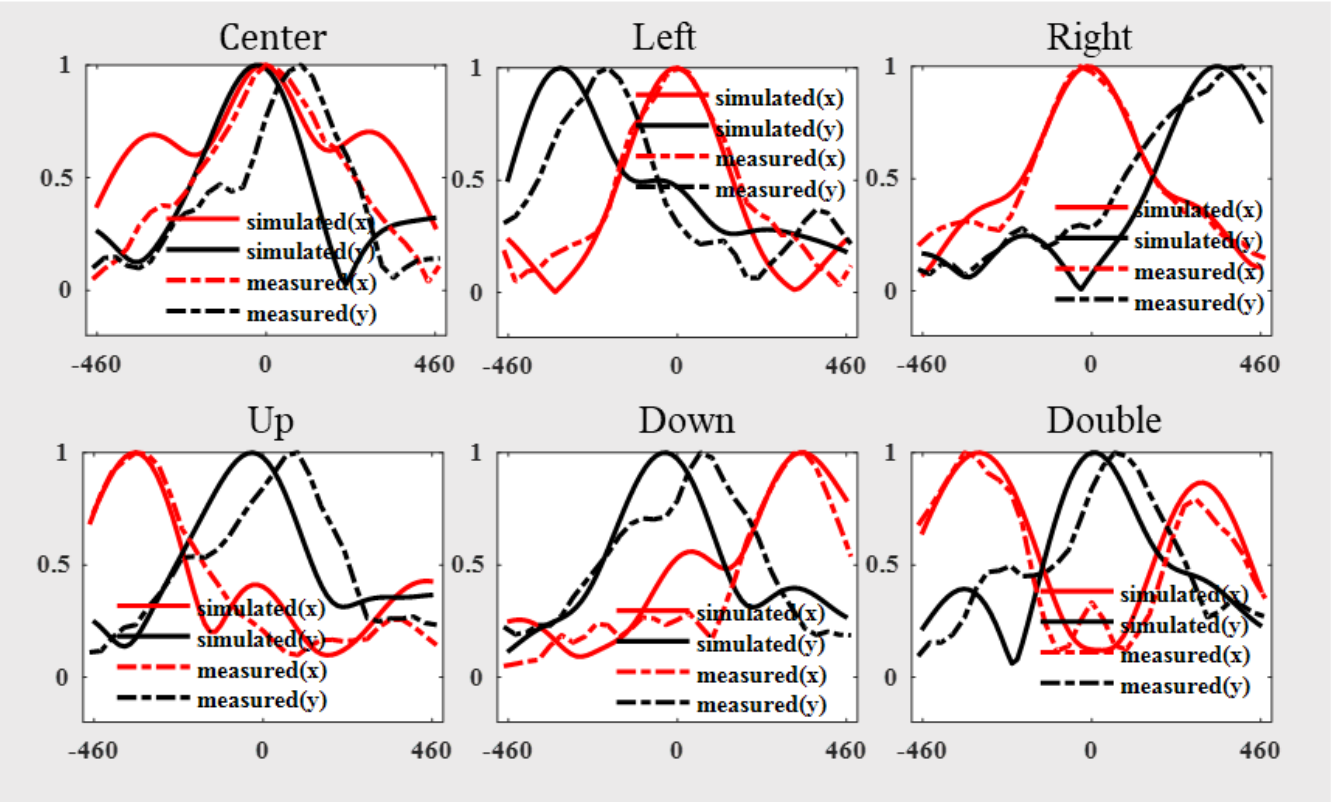

Fig. 7. The normalized simulated and measured intensity distributions along the vertical ($x$/mm) and horizontal ($y$/mm) lines at the maximum. The simulated and measured results are respectively represented by solid and dotted lines. The results along the $x$ and $y$ direction are respectively plotted in red and black fonts.

In addition, in order to evaluate quantitatively the accuracy of proposed methods in Section III.A and B, the normalized measured and predicted intensity distributions, along the vertical ($x$ direction) and horizontal ($y$ direction) lines at the maximum (represented by white dotted lines in Fig. 5), are depicted in Fig. 7. The corresponding full-widths at half-maximum (FWHM) are summarized in Table V. Overall, the experimental results are well consistent with the theoretical predictions. However, there are some notable discrepancies between the experimental results and numerical predictions probably due to complicated indoor measurement environment and other possible effecting factors. For instance, the measured focal points are diverged from the predicted locations along $y$ direction, and some of the measured FWHM are much larger than predicted results. One possible reason for the discrepancy in focused position is due to the uncontrollable deviations from the location of Wi-Fi antenna along the $y$ direction in actual experiments. The actual focal points will diverge along with the source, since the received field is derived from phase distribution on metasurface, wherein the phase distribution is closely related to the position of Wi-Fi antenna. Moreover, the mirror reflection will also cause deviations in the $y$ direction of the focal points, which can be obviously observed from the cases of “Center” and “Right”. As for the larger FWHM, they may be caused by mirror reflection and complex experimental environment since the measured results are obtained in an office room. Despite of the discrepancy, the measured results have satisfied our demand of customizing diverse spatial intensity allocation of mimic Wi-Fi signals.

Then, the important efficiency analysis has also been conducted in this subsection. Here, we define the focus efficiency as the ratio of $P_f$ (the power of the focus area in the designated observation plane) and $P_t$ (the total power in the designated observation plane). Two kinds of focus efficiency corresponding to the aforementioned six simulated cases observed in z=0.984m plane have been considered, in which the $P_f$ is obtained within two kinds of focus regions where the normalized field intensity is larger than -3dB and -6dB, respectively. The results are summarized in TABLE II.

TABLE II
The FOCUS EFFICIENCY FOR SIX DIFFERENT CASES

| Cases | Center | Left | Right | Up | Down | Double |
|---|---|---|---|---|---|---|
| -3dB | 25.36% | 30.73% | 40.24% | 28.51% | 35.27% | 21.27% |
| -6dB | 37.42% | 45.19% | 59.32% | 41.81% | 51.96% | 38.65% |

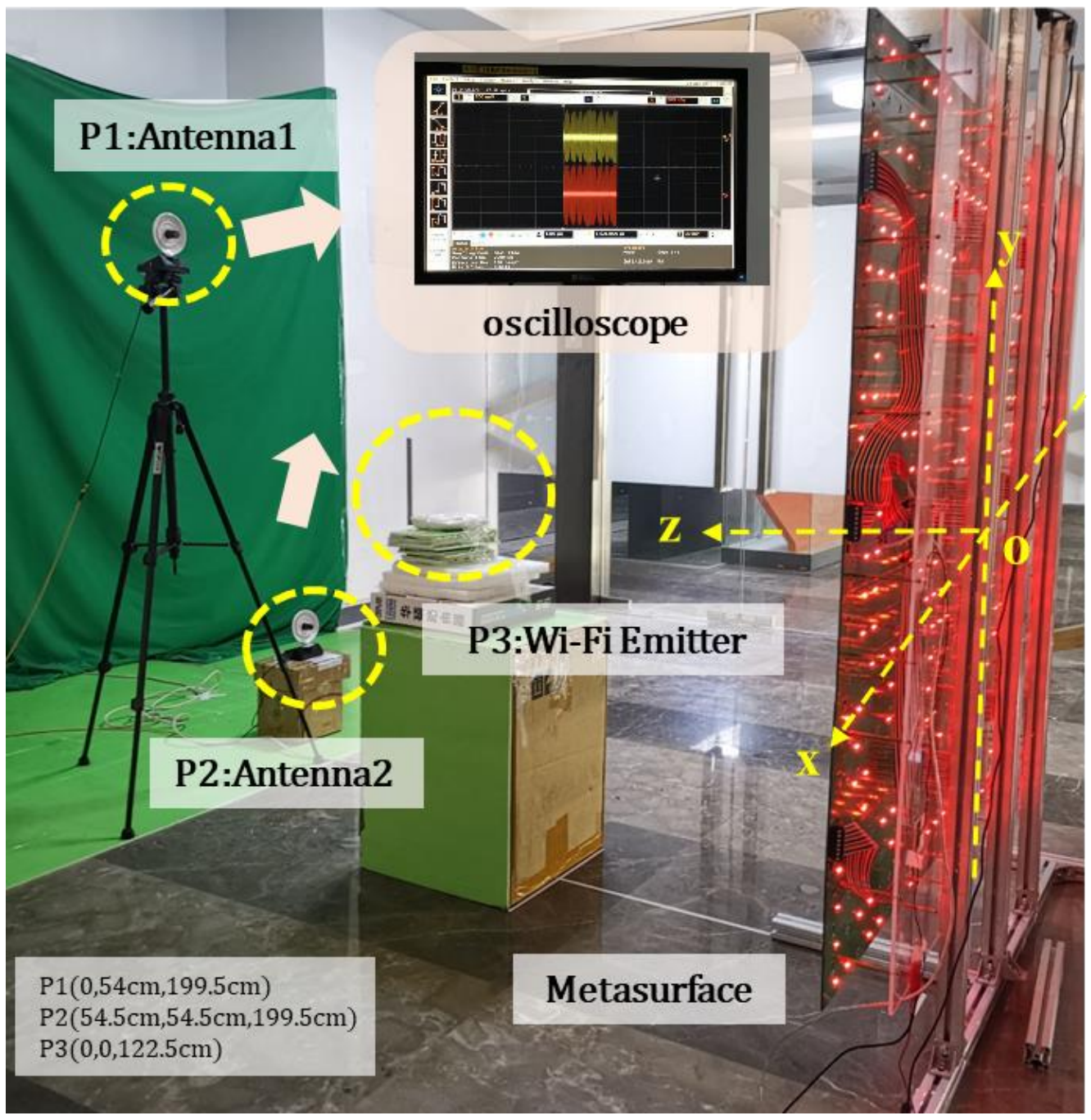

Fig. 8. Practical experimental setup using the commercial Wi-Fi device.

## IV. Experiments Against Commodity Wi-Fi Signals

In this section, selected experimental results are provided to demonstrate the performance of the proposed strategy for reallocating the spatial energy of commodity Wi-Fi signals. With reference to Fig. 8, our experimental setup consists of four building components: a programmable coding metasurface, a commercial 802.11n Wi-Fi router (Mercury MW150R), a pair of antennas, and an oscilloscope (Agilent MSO9404A). The Wi-Fi router, working at the 7$^{th}$ operational channel, i.e., 2.431~2.453 GHz, is randomly placed somewhere in front of metasurface, e.g., P3(0, 0, 1.225m). Two commercial parabolic antennas connected with two ports of the oscilloscope are deployed to acquire the commodity Wi-Fi signals bounced off the metasurface. Moreover, the oscilloscope is set with the sampling rate of 10GHz and the trigger level of 0.02V.

### *A. Focusing of commodity Wi-Fi signals at a single location*

As the first set of experimental investigations, we examine the performance of the designed metasurface and associated optimization algorithm in focusing the Wi-Fi signals at a single local location. In our experiments, the desired focus is located at P1 (0, 0.540m, 1.995m). Five different coding sequences of metasurface are considered, i.e., the full dark, full bright, two random coding sequences, and an optimized coding sequence. Here, we denote the full dark by setting all PIN diodes of metasurface to be OFF status, while the full bright by setting all PIN diodes to be ON status. The acquired time-domain Wi-Fi signals within the duration of 2us at P1 are presented in the second row of Fig.9. Moreover, the average power of Wi-Fi signals, corresponding to the aforementioned five coding sequences of metasurface, has been reported in Table III. It is clear that the intensities of received Wi-Fi signals vary significantly with the choices of coding sequences of metasurface, and the signal with the remarkably enhanced intensity can be achieved by a factor of around 10 compared with the case of "full bright" if the programmable metasurface is written with the optimized coding sequence. In addition to the enhanced signals, from this set of results, we can also conclude that the spatial intensity of Wi-Fi signals at given position can be controllably diminished using the proposed strategy.

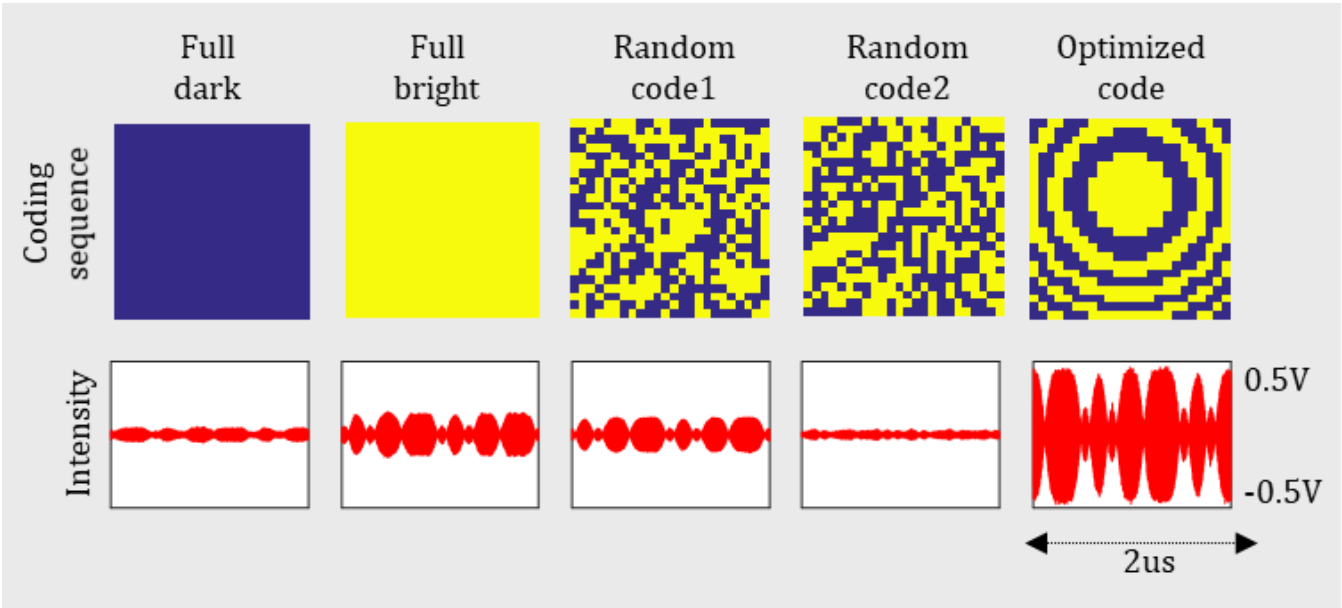

Fig. 9. Experimental results of commercial Wi-Fi signals at location P1 for five different coding sequences of metasurface.

TABLE III
AVERAGE POWER AT SINGLE LOCATION FOR DIFFERENT CODING SEQUENCES

| Code | Full dark | Full bright | Random Code1 | Random Code2 | Optimized code |
|---|---|---|---|---|---|
| Average power/a.u. | 6.55e-4 | 6.70e-3 | 4.60e-3 | 2.45e-4 | 7.01e-2 |

### *B. Focusing of commodity Wi-Fi signals at multiple locations*

Here, we will further examine the performance of the designed programmable coding metasurface and associated optimization algorithm for focusing the commodity Wi-Fi signals at multiple locations, i.e., P1 (0, 0.54m, 1.995m) and P2(-0.545m, 0.545m, 1.995m), as marked in Fig. 8. We consider three cases, i.e., focusing at a single position P1, focusing at a single position P2, and focusing simultaneously at two positions P1 and P2. By performing our optimization algorithm outlined in Section III. B, we can obtain the coding sequences of programmable coding metasurface corresponding to aforementioned three scenarios, as shown in the first row of Fig. 10. Correspondingly, the distributions of the predicted normalized intensity of Wi-Fi signals have been plotted in second row of Fig. 10. The time-domain Wi-Fi signals within the duration of 2us, captured by two receiving antennas at P1 and P2, are depicted in the third and fourth rows of Fig. 10. Apparently, the experimental results are in good agreement with our numerical predictions. The average power at both positions for these optimized coding sequences are summarized in Table IV. As indicated, the average power of measured Wi-Fi signals at the focused point P1 for the first coding sequence is larger than the one at focused point P2 for the second coding sequence. This does make sense since the location P2 is more far away from the Wi-Fi router and metasurface than that of P1. In this way, we can say that our method can support the energy allocation of commodity Wi-Fi signals in a controllable way.

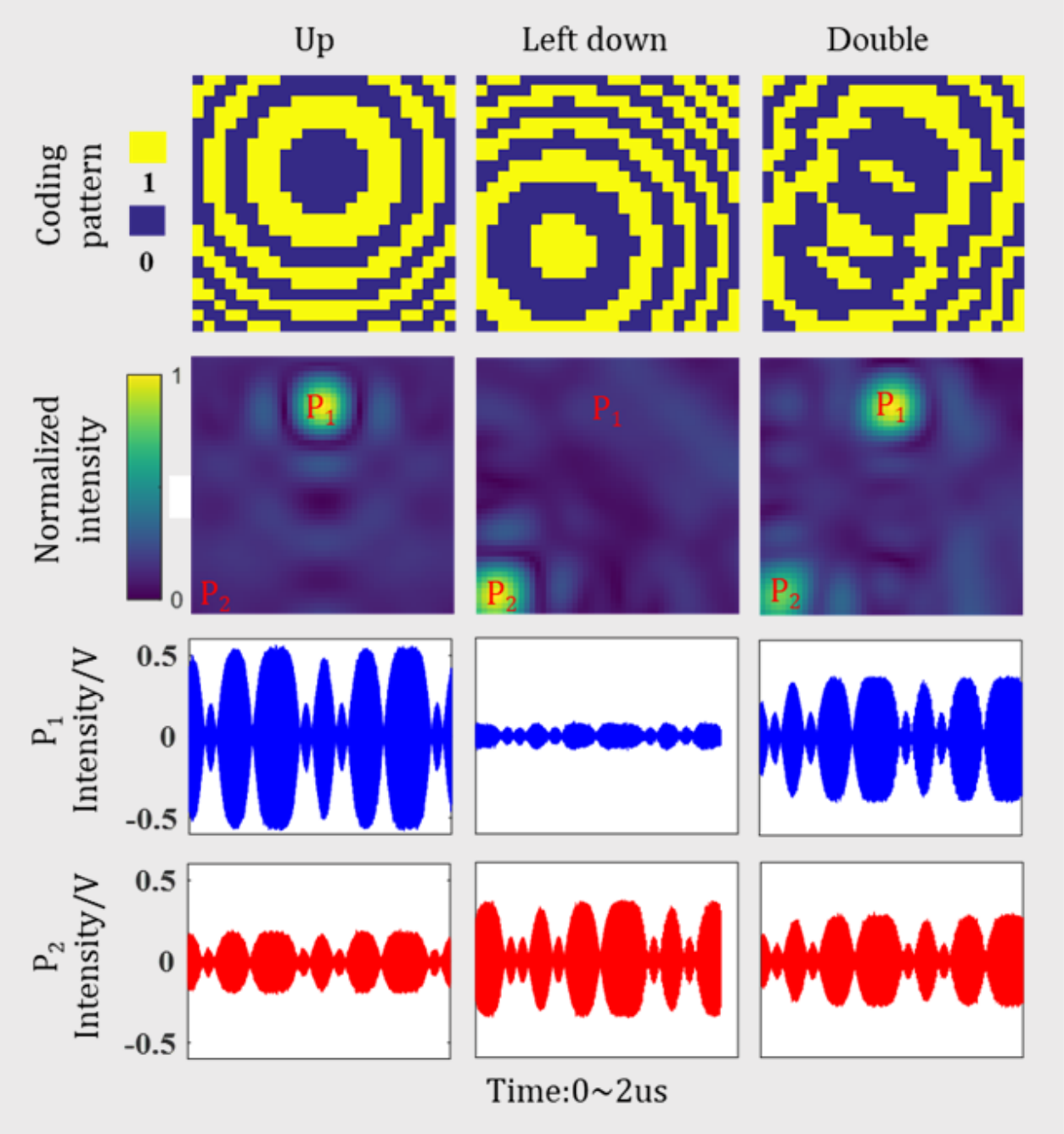

Fig. 10. Experimental results for the focusing of commodity Wi-Fi signals at multiple locations. The first row corresponds to three optimized coding sequences at the distance of z=1.995m plane. The second row shows the corresponding normalized spatial intensity distributions predicted at the z=1.995m plane using the proposed method. The third to fourth rows are the captured commercial Wi-Fi signals at different locations P1 and P2, respectively.

TABLE IV

AVERAGE POWER AT MULTIPLE LOCATIONS FOR OPTIMZIED CODING SEQUENCES

| Code | Up | Left down | Double |
|---|---|---|---|
| P1:average power/a.u. | 7.92e-2 | 1.20e-3 | 3.66e-2 |
| P2:average power/a.u. | 8.20e-3 | 2.87e-2 | 1.87e-2 |

Before concluding this section, we also would like to say that instead of point-like energy enhancements of Wi-Fi signals, other more complicated energy allocation of Wi-Fi signals also can be achieved in principle by adopting the proposed strategy. In order to show this, another set of numerical simulations are conducted, and corresponding results are provided in Fig. 11. Here, we consider such a scenario in which the metasurface is mounted on the ceiling inside a house. Then the indoor Wi-Fi signals are controlled to be concentrated in some desired areas simultaneously such as two bedrooms and a table in the living room, as shown in Fig. 11(a-b), by properly re-engineering the coding sequences (as depicted in Fig. 11(c).

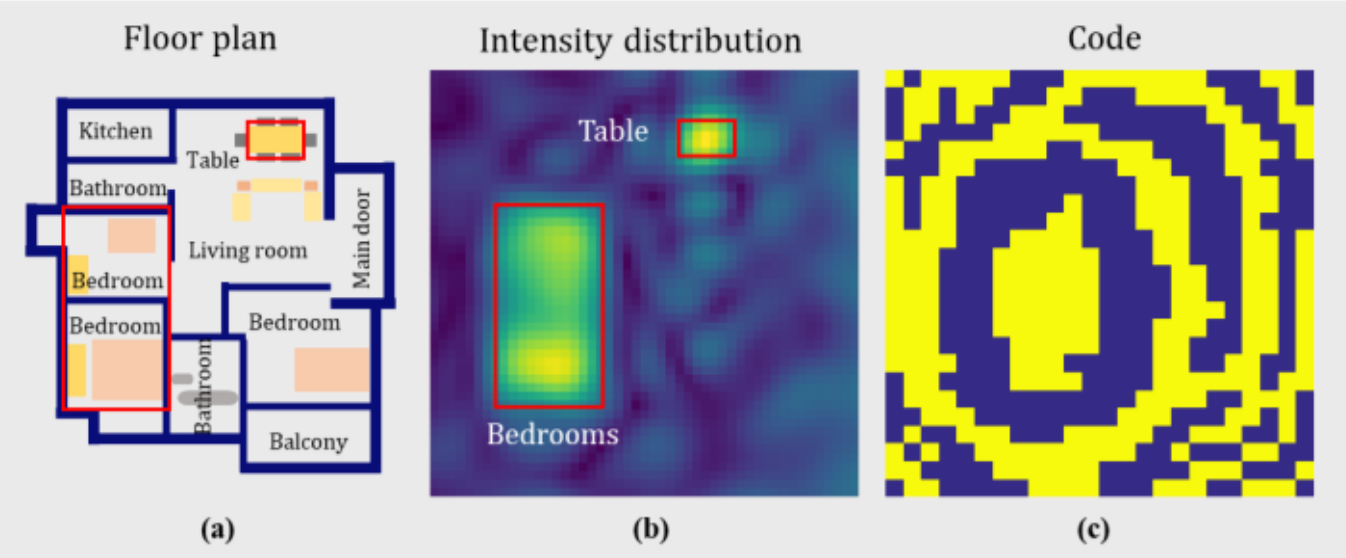

Fig. 11. The simulation of complicated energy allocation of Wi-Fi signals in a typical indoor environment. (a) The floor plan of a common house. (b) The simulated energy distribution in which the Wi-Fi signals are controllably reallocated to be concentrated in two bedrooms and on a table simultaneously. (c) The corresponding coding sequence of metasurface.

## V. CONCLUSION

In this work, we have developed an efficient approach to manipulate commodity Wi-Fi signals in a flexible, dynamic and inexpensive manner. To this end, we design, fabricate and test an electronically-controllable phase-binary coding metasurface, which works at around 2.4GHz. Besides, we have developed an efficient optimization algorithm to allocate the spatial intensity of commodity Wi-Fi signals in real time. Experimental results have been provided to show that our strategy works very well for controlling commodity Wi-Fi signals of IEEE 802.11n protocol in a flexible and dynamic manner.

Our method could find potential applications in several practical scenarios. For instance, the developed strategy can be used to enhance or diminish the received Wi-Fi signals at intended destination, which may have huge potentials in developing the next-generation wireless communication system and resolving well-concerned problems of the path loss and multi–path fading effects in the conventional wireless communication. Moreover, since the spatial Wi-Fi signals can be arbitrarily manipulated by control commands sent from master computer through a real-time communication, our proposed strategy can be utilized to build an intelligent Wi-Fi environment in the future. The Intelligent Wi-Fi environment, envisioned as a reconfigurable space, can programmatically control wireless signals following software directives, which will play a prominent role in distributing and processing information on physical layer. To summarize, it could be expected that our strategy will pave promising avenues for future wireless networks and smart home.

TABLE V

The FWHMs along the vertical and horizontal lines at the observation plane of z=1.531m

| *Type* | *Center* | | *Left* | | *Right* | | *Up* | | *Down* | |
|---|---|---|---|---|---|---|---|---|---|---|
| *FWHM/mm* | $D_x$ | $D_y$ | $D_x$ | $D_y$ | $D_x$ | $D_y$ | $D_x$ | $D_y$ | $D_x$ | $D_y$ |
| *Predicted* | 247.11 | 227.20 | 224.48 | 220.86 | 238.10 | 262.40 | 226.40 | 255.24 | 228.20 | 236.22 |
| *Measured* | 254.96 | 197.68 | 248.49 | 235.56 | 240.10 | 471.20 | 237.40 | 239.25 | 218.93 | 291.91 |

| *Type* | Double | | | |
|---|---|---|---|---|
| *FWHM/mm* | $D_x^u$ | $D_y^u$ | $D_x^d$ | $D_y^d$ |
| *Predicted* | 262.5 | 224.5 | 216.3 | 215.44 |
| *Measured* | 265.1 | 222.62 | 197.7 | 203.23 |

## ACKNOWLEDGMENT

This work is supported by the National Key Research and Development Program of China (Grant No. 2017YFA0700203).